# Big data analytics architecture—an application in manufacturing systems

Mahdi Fahmideh, Ghassan Beydoun

Faculty of Engineering and Information Technology, University of Technology Sydney, Sydney 2007, Australia

**Abstract**

**Context:** The rapid prevalence and potential impact of big data analytics platforms have sparked an interest amongst different practitioners and academia. Manufacturing organisations are particularly well suited to benefit from data analytics platforms in their entire product lifecycle management for intelligent information processing, performing manufacturing activities, and creating value chains. This requires re-architecting their manufacturing legacy information systems to get integrated with contemporary data analytics platforms. A systematic re-architecting approach is required incorporating careful and thorough evaluation of goals for data analytics adoption. Furthermore, ameliorating the uncertainty of the impact the new big data architecture on system quality goals is needed to avoid cost blowout in implementation and testing phases.

**Objective:** We propose an approach to reason about goals, obstacles, and to select suitable big data solution architecture that satisfy quality goal preferences and constraints of stakeholders at the presence of the decision outcome uncertainty. The approach will highlight situations that may impede the goals. They will be assessed and resolved to generate complete requirements of an architectural solution.

**Method:** The approach employs *goal-oriented modelling* to identify obstacles causing quality goal failure and their corresponding resolution tactics. It combines *fuzzy logic* to explore uncertainties in solution architectures and to find an optimal set of architectural decisions for the big data enablement process of manufacturing systems.

**Result:** The approach brings two innovations to the state of the art of big data analytics platform adoption in manufacturing systems: (i) A systematic goal-oriented modelling for exploring goals and obstacles in integrating manufacturing systems with data analytics platforms at the requirement level and (ii) A systematic analysis of the architectural decisions under uncertainty incorporating stakeholders' preferences. The efficacy of the approach is illustrated with a scenario of reengineering a hyper-connected manufacturing collaboration system to a new big data architecture.

## 1 Introduction

Product lifecycle management is a data intensive process comprising market analysis, product design, development, manufacturing, distribution, post-sale, and recycling (Stark, 2015). The process involves a variety of voluminous data, e.g. customers' comments on social media, product functions, product configuration, and failure incidences reported by installed sensors to monitor parameters of environment and products. Manufacturing organisations view such data as a valuable business asset to achieve good performance and to reduce cost in the product lifecycle. They also regularly seek to increase their productivity using new advanced information technologies that place further demand on their data processing storage requirements such as Internet of Thing (IoT) and radio-frequency identification (RFID) tags in their daily production. For example, Toyota automotive company equip cars with smart sensors and continuously collecting data about its locks, location, ignitions, and tyres which can be later used by the manufacturer assembly. Continuous product innovations lead to further product data generation coupled with a great diversity of types, sources, meaning, and format.

Given its increasing volume and variety, manufacturing data is increasingly difficult to process using common manufacturing data platforms be they computer aided design (CAD), supply chain management (SCM) manufacturing execution system (MES), or enterprise resource planning (ERP). Indeed, the high volume, velocity, variety, veracity, and value adding data requirement all point to the need to complement manufacturing systems with *big data* platforms (McAfee, Brynjolfsson, & Davenport, 2012). New platforms such as Apache Hadoop, Google's Dremel, or S4 are promising ways forward to address the abovementioned processing complexity (Lycett, 2013). They provide a support for capturing, processing, and visualising large volume of data sets that organisational systems may have collected over the years.

Taming big data has the potential added benefit to analyse real-time data across different phases of product lifecycle management from receipt of a customer's order, identifying promising customers, collecting variable data about the quality of raw material, selecting a detailed design, procuring, selecting suppliers and outsourcing policies, and product warehousing, maintenance, recycling, and to identifying labor errors have been discussed (Li, Tao, Cheng, & Zhao, 2015), (Protiviti, 2017), (Waller & Fawcett, 2013), (Bi & Cochran, 2014), (Dubey, Gunasekaran, Childe, Wamba, & Papadopoulos, 2016).

Compared to others fields such as electronic commerce, financial trading, health care, and telecommunication, the manufacturing field seems to be slow in pace in adoption big data analytics platforms in their business processes (Li et al., 2015). This could be attributed to the high capital costs associated with manufacturing systems which automate, process, and integrate data flows between one or more above phases and typically composed of a number of software systems, machines, transportation devices, and so on (Camarinha-Matos, Afsarmanesh, Galeano, & Molina, 2009). Nevertheless, manufacturing organisations recognise the value of big data analytics and the fact that their adoption failure poses a risk to their operating and financial performance (Protiviti, 2017). Gartner reports that 60 percent of big data projects fail to get piloting and production due to reasons such a lack of adequate IT skill set, inability to understand stakeholders requirements in utilising data analytics, and disparate legacy systems (Gartner, 2015), (Wegener, 2013). Thence, reluctance of manufactures in moving to these platforms is unsurprising. Some are also still figuring out what kind of data is worthy for advanced data analytics and which stage of product lifecycle management is suitable to utilise big data analytics platforms (Govindarajan, Ferrer, Xu, Nieto, & Lastra, 2016), (S. Jha, Jha, O'Brien, & Wells, 2014), (Bi & Cochran, 2014).

It has been a long-standing acknowledgement that a poor system upgrade with a new technology can have far reaching consequences in later stages that are costly to rectify. This continues to be permeating theme in adoptions of big data analytics platforms in manufacturing systems. Particularly, these may involve many competing goals, e.g. security, performance, reliability, scalability, maintainability, and the development cost. There are also unforeseen risks. As articulated by Protiviti: *"Manufacturers should have clear and easily definable goals. As a part of that planning process, companies need to determine whether the systems they have in place will achieve the desired results and/or what enhancements might be required"* (Protiviti, 2017). Manufacturing also tend to have their own goals and preferred competitive dimensions with respect to taking advantages of data analytics platforms to augment their systems (Wang, Xu, Fujita, & Liu, 2016). A system architect, who is responsible to the design high-level big data enabled solution architecture, should meticulously specify goals of multiple stakeholders, analyse potential risks, and make a right balance among operationalisation of the goals in adopting these technologies (Lee, Kao, & Yang, 2014) (Wang et al., 2016). For example, using a poor big data visualisation technology may negatively affect performance, scalability, and real-time data processing coming from sensors. The choice of a data mining algorithm to process sensor data across the product line may also impact the real-time performance of control systems. An early stage analysis of big data adoption goals gives an opportunity in exploring countermeasures to tackle probable risks in advance rather than drowning in narrow aspects of these technologies.

Furthermore, uncertainties about the impact of decisions on data analytics adoption goals are unavoidable, as in any other adoption endeavors. That is, a lack of complete knowledge about the actual consequences of architectural decisions is a fact. For instance, the raw feedback data generated by online customers about produced cars that are processed by data analytics platforms may produce some uncertainties in terms of the interpretation of data. The choice of a data visualisation technique may have an uncertain impact on the reliability of other generated diagnostic reports about a product due to inherent uncertainties of data sources. On the other hand, the system architect is still expected to make right choices in such uncertain circumstances. The quest for a risk-aware early stage analysis of big data adoption in making critical and uncertain decisions remains a top priority as highlighted in (Pal, Meher, & Skowron, 2015) and (C. P. Chen & Zhang, 2014).

This paper provides an approach aiding system architects for goal-obstacle analysis of big data solution architectures and selecting architectural decisions using imperfect information. It provides a step-by-step goal-obstacle analysis process to address uncertain risks. It then produces a complete set of requirements, ranks candidate architectures based on the fuzzy logic, and ultimately to find an optimum architecture. The contributions of the paper are thus two folds: (i) providing a goal-oriented approach for reasoning about architectural requirements at the early stage of adoption (ii) dealing with uncertainty about the impact of integrating data analytics platforms on manufacturing systems which can be unforeseen at the requirement time. It should be noted that the approach is applicable outside the context of manufacturing. The emphasis of the validation and the exemplars are manufacturing, however the goal modelling and obstacle resolution approach can be easily applied to other contexts (Fahmideh & Beydoun, 2018).

The rest of this article is organised as follows: Section 2 provides the background of this study including a motivating scenario and fundamental concepts used in the proposed approach. Section 3 details the approach. Section 4 illustrates the approach in an exemplar scenario of re-architecting a legacy car manufacturing system to utilise data analytics platforms. Section 5 outlines other related studies. Finally, Section 6 summarises the article with a discussion of limitations and future research directions.

## 2 Background

### 2.1 A motivating scenario of moving manufacturing systems to data analytics platforms

We adopt and extend an exemplar reengineering scenario of a hyper-connected manufacturing collaboration system (HMCS) (Lin, Harding, & Chen, 2016). HMCS provides a platform to enable several partners of Toyota car manufacturing to collaborate and to share knowledge about car products and parts across the product line. At the core of the HMCS architecture, a data Extract-Load-Transform (ELT) processes and integrates data streams from multiple manufacturing parties and different types of databases including those storing data coming from sensors in the product line or from buyers. The process has a middleware layer that includes rules and logic for mapping data between different formats. An additional data stream comes from the online Toyota buyer conversations that appear in social media, such as Twitter, posts, Internet server logs, and blogs. This stream provides feedback on recent purchase experiences, warranty claims, repair orders, service reports and others which can be used to uncover actionable trends and to generate appropriate early-warning signals to the manufacturing process. The stream is increasingly voluminous and it generates millions of items per day.

The heterogeneity of data sources and the large volume of data strain the ETL. It often becomes a bottleneck and incapable of identifying all patterns and generating statistical reports. To resolve this, the IT department of Toyota aims to deploy multiple data analytics platforms. The aim is to enable extraction and management of both sources of data, i.e. internal manufacture parties and the unstructured data in the online Toyota buyer conversations. A system architect is appointed specifically to design a solution to upgrade and integrate the ETL with services offered by the data analytics platforms. An immediate concern of the architect is a cost benefit analysis of the adoption of the platforms evaluating the risks and mapping the way forward. The following questions are pertinent to the system architecture: What are ETL system quality goals? How these may be positively or negatively affected if data analytics platforms are utilised? Will higher ETL performance be attainable in all circumstances? What obstacles are likely to occur during and after re-engineering

ETL to data analytics platforms and what are their severity? What countermeasures can be added in advanced to negate such obstacles and do they have any side effects? Answering these questions is a challenging task as it involves reasoning with a long chains of 'what-if' scenarios and their uncertain impacts on goals given by the stakeholders of HMCS. Additionally, these impacts are even imprecise and may be hard to quantify. Human judgments are often too vague for using exact numerical values, let alone in an early stage of a solution architecture design.

Towards designing a big data solution architecture, we offer an approach that explicitly relates manufacturing system high-level quality goals to potential obstacles, highlights architectural requirements in addressing them and assesses their impact on stakeholders' goals, calculates the uncertainty of the various impacts, and finally shortlists candidate architectures satisfying the goals.

## 2.2 Goal-oriented requirement modeling

Goal-oriented reasoning approaches such as KAOS and i* are means for the elicitation, elaboration, and analysis of system requirements (Yu & Mylopoulos, 1994). We choose KAOS (Keep All Objects Satisfied) modelling framework. It defines two components: (i) a modeling language including concepts such as *goal*, *obstacle*, *agent*, *operation*, and *domain objects* and (ii) a method specifying a series of steps to elaborate and analyse goals (Van Lamsweerde, 2009). In this article, we only use to concepts goal and obstacles as defined in the following.

A goal is a prescriptive statement of intention that a system should satisfy through the cooperation of agents. A goal has a name and a specification expressed using natural or formal languages. The specification defines what the goal means and its satisfaction conditions. Goals may range from high-level business objectives to fine-grained technical ones. All goals are continuously refined into sub-goals until all sub-goals can be assigned to a single agent, i.e. a user or a system component. In this article, we refer to common system quality goals such as performance, security, and maintainability. The method part of the KAOS framework provides a process to create a goal model through a hierarchical refinement process.

A goal which is stated without considerations of unexpected conditions in a real environment that may cause their failures is considered an optimistic goal (Letier, 2001; van Lamsweerde & Letier, 2000). Taking a more realistic and a deeper look, it is prudent to consider these conditions and to construe them as obstacles. Obstacles are duals of goals in the sense that as goals represent desired conditions, obstacles represent undesirable conditions (Letier, 2001) that should be systematically and concomitantly identified. They need to be assessed and tackled via defining resolution tactics at an early stage of a system development to identify any needs to modify the goals (Letier, 2001). As such, in KAOS, resolving obstacles includes steps for generating and selecting alternatives to resolve obstacles. Selection of a subset of decision alternatives satisfying goals faces a multi-criteria decision making problem (MCDM) with the possibility of different priority of goals in a view of stakeholders needs. Any uncertainty in selecting options may also occur in terms of a range of impacts that options may have on goals. It is often the case that stakeholders may express such impacts qualitatively or using imprecise measures because their judgements are unavoidably vague and indescribable with exact numerical values. For example, the impact of choosing a data mining algorithm for processing data coming from sensors might be expressed in linguistic terms or a range of values rather than a crisp and single number. Subsequently, each decision alternative may fall within a range. Comparing two decision alternatives with overlapping in their impact on goals is not easy. There is often a need to consider trade-offs amongst various alternatives. Whilst MCDM frameworks can help in comparing, prioritising, and selecting the most suitable resolution tactics, they do not reflect the uncertainty in human thinking style. Fuzzy logic can better handle the uncertainty. For instance, expressions such high performance or low cost become usable. We make a synergy between the KAOS approach and fuzzy logic to cope with various sources of uncertainties in integrating manufacturing systems with data analytics platforms.

## 2.3 Fuzzy set theory

Fuzzy set theory analogises human judgment at the presence of proximate information and uncertainty in decision making (Zadeh, Fu, & Tanaka, 2014). Whilst classic sets define crisp values, fuzzy sets shows groups of data with boundaries that are not crisp. This provides a better capability to resolve real-world problems, which unavoidably involve imprecise and noisy parameters. Accordingly, linguistic expression of variables is the central aspect of fuzzy logic where general terms such a *very large*, *large*, *medium*, *small*, *too small* are used to represent a range of numerical values.

The fuzzy set, originally proposed by Zadeh, is defined as follows (Zadeh et al., 2014): In a universe of discourse U, a fuzzy subset A is characterised by a membership function F where each member of x ∈ U is associated with a number of F in the internal [0,1], denoting the membership of x in A. The impact linearly decreases from very low to very high. This range of impact is represented using a triangular fuzzy value (Pedrycz, 1994). For example, the choice of a particular data mining algorithm for processing sensor data may have *Very High* positive impact on the overall system performance. Such values may be available from statistical data of similar architecture designs in other systems, system architect's experience, or expert judgment.

The ability to quantitatively analyse alternatives in a big data solution architecture manufacturing systems can be achieved via representing uncertain parameters as fuzzy numbers which belong to fuzzy sets. Instead of showing the anticipated impact of an architectural decision alternative on system quality goals as a discrete point, we represent such impact as a range of values. This is more aligned with human judgment in conceptualisation and representation of uncertainty.

## 3 The approach

Our approach, as shown Figure 1, has two steps: In the first step, high-level goals of adopting data analytics platforms are identified. Their operational alternatives and potential corresponding obstacles are then identified and assessed. Obstacles that are deemed to be severe are resolved through generating resolution tactics. Step 1 engages the stakeholders and iterates over the sequence of refinements akin to the one described in (Lim & Finkelstein, 2012). In the second step of the approach, the impact on data analytics adoption goals is analysed to optimise the chances of success in goal achievement. The overall output of the approach as shown in Figure 1 is a set of solution architectures, ranked on the basis of likelihood of satisfying specified quality goals. One final selected architecture from this set gets later incorporated to reengineer the existing manufacturing systems to data analytics platforms. To illustrate the details the steps, the scenario presented in Section 2.1 is used as an exemplar in what follows.

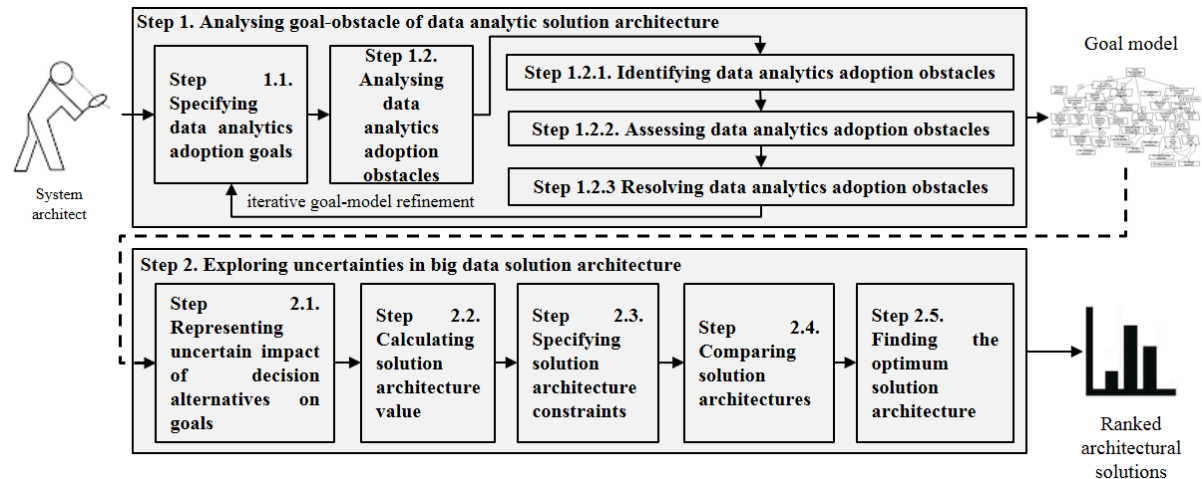

Figure 1. Proposed approach

### 3.1 Step 1. Analysing goal-obstacle of data analytics solution architecture

The first step is based on KAOS modelling framework and uses the notation shown in Table 1. Step 1 includes these two sub-steps:

(i) *Step 1.1. Specifying data analytics adoption goals* targeting by data analytics platforms and possibly decision alternatives to operationalise the goals,

(ii) *Step 1.2. Analysing data analytics adoption obstacles* for identifying goal failure causes, assessing their likelihood and criticality of consequence, defining resolution tactics and decision alternatives. This is a mitigation step aiming at reducing the likelihood of the obstacles occurring or eliminating it altogether.

Table 1. Notations used for the goal modelling

| Modelling element | Definition | Graphical notation |
|---|---|---|
| Root (overall goal) | The overall goal of adopting data analytics platforms contributing to systems. | |
| Goal | A quality goal that is expected to be satisfied via utilising data analytics platforms. | |
| Obstacle | A technical/none-technical exceptional situation/condition preventing the goal satisfaction. | |
| Architectural decision | A generic architectural solution either to operationalise a goal or to tackle an obstacle. | |
| Architectural decision alternative | A technique, tool, or technology taking to operationalise an architectural decision. | |

Like many IT projects that are inherently conducted cooperatively by a team, the rationale for embedding modeling within the approach is to broaden stakeholder participation (e.g. system architect, developers, and users) in the entire goal analysis. This will enable appropriate documentation and argumentation surrounding goals, obstacles, architectural decision alternatives and to coordinate the design effort, and to ensure convergence to potential architectures. The following subsections provide technical details of Step 1.

**Step 1.1 Specifying data analytics adoption goals.** Seven goals are set for the integrating ETL with data analytics platforms (Figure 2): *g1.Achieve [Processed social media weekly under expected time]*, *g2.Achieve [Processed sensor data under expected time]*, *g3.Achieve [Improved availability]*, *g4.Achieve [Maintained interoperability with other big data platforms]*, *g5.Achieve [Improved data visualisation]*, *g6.Achieve [Maintained data security on big data platforms]*, and *g7.Achieve [Increased unstructured data storage capacity]*.

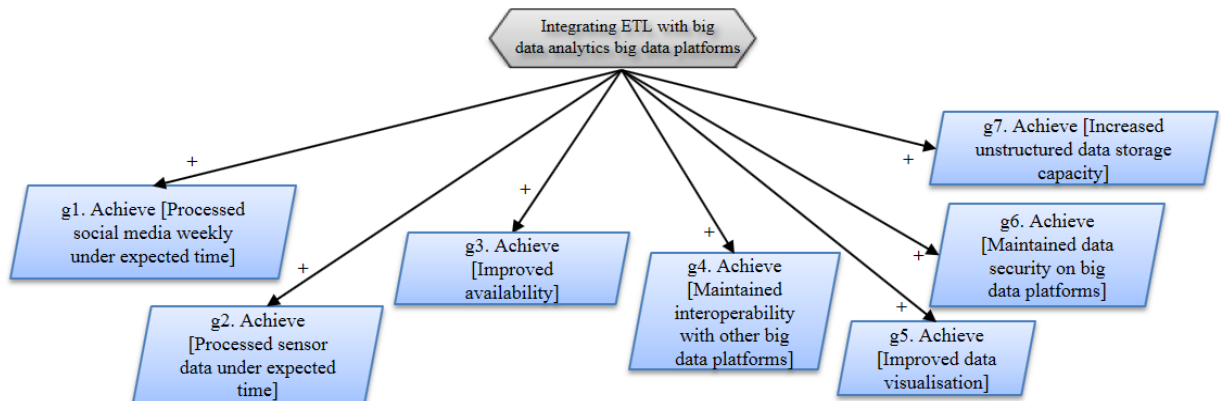

Figure 2. Goals of integrating ETL with data analytics platforms

ETL databases is rapidly growing in size and reaching several terabits of data collected by sensors in the product line. An unlimited data storage capacity is required. The system architect documents the specification of goal *g7.Achieve [Increased unstructured data storage capacity]* as follows:

    **Goal** g7.Achieve [Increased unstructured data storage capacity]
    **Category** scalability goal
    **Definition** Batches of data records from manufacture product line provided by installed sensors should be captured and stored continuously. These records consist of data monitored by robots about assembling Toyota parts in the production line.
    **Quality Variable** storageSize: Batch → Size

**Definition** The required capacity in storing records of data collected by sensors in a working day.
**Sample Space** The set of daily cars is assembled and delivered to the end of the product line.
**Objective Functions** At least one gigabyte of records (e.g. images, events, logs, and errors) are generated at the end of a working day. The database should be able to store this volume.

For the goal *g1.Achieve [Processed social media weekly under expected time],* the following definition is documented:

**Goal** g1.Achieve [Processed social media weekly under expected time]
**Category** performance goal
**Definition** Relevant data of buyers experience should be collected from Twitter, posts, Internet server logs, and blogs and then be processed every week and results be available on Monday at 12am. This data from customers can be in the form of feedback or complaints, maintenance requests, part orders, or product comparisons.
**Quality Variable** ProcessedTime: Batch → Time
**Definition** The required capacity to storage sensor data at the end of the day.
**Sample Space** The set of daily cars that are assembled on the product line and delivered.
**Objective Functions** The processing of all the collected data and generating reports should be started from 12 am on Saturday and finished at midnight on Sunday.

Goals are operationalised through different architectural decisions, i.e. tactics, techniques, and technologies. As shown in Figure 3, to realise the goal *g7.Achieve [Increased unstructured data storage capacity],* the system architect considers five mainstream big data storages namely *a17.MongoDB*, *a18.Accumulo*, *a19.HBase*, *a20.Cloudant*, and *a21.BigTable*. Likewise, to implement the goal *g1.Achieve [Processed social media weekly under expected time]*, both technologies *d1.scheduler* and *d2.social media data processing* can be used where each has different alternatives to be employed (Figure 3).

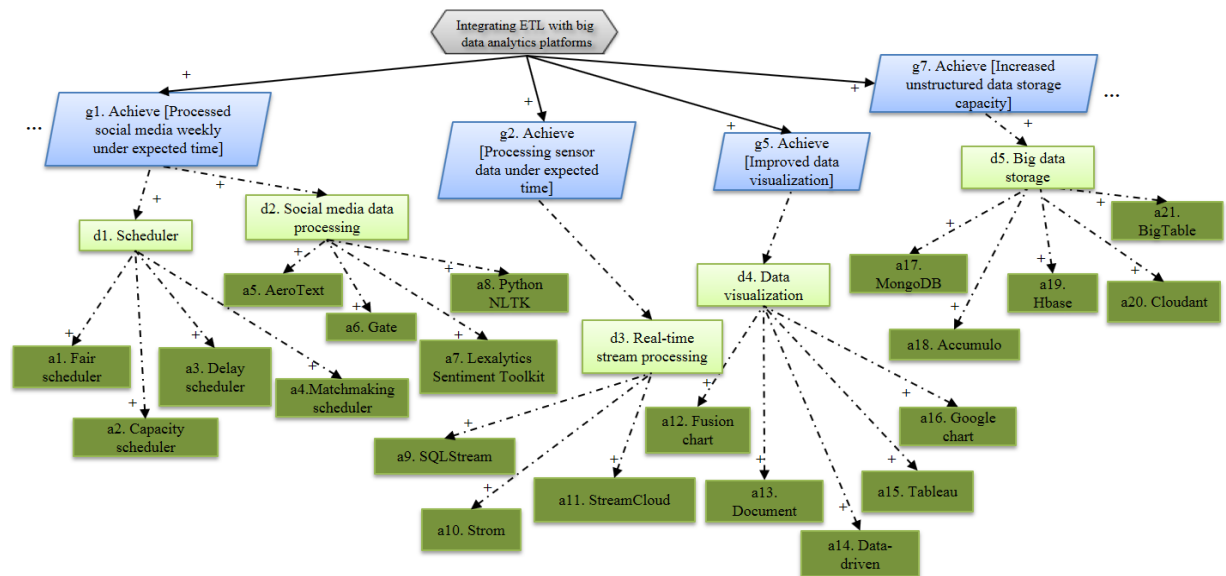

Figure 3. Operationalisation of the goals through architectural decisions and related implementations

**Step 1.2. Analysing data analytics adoption obstacles.** Normally, goals neglect unexpected situations that may cause their failures in operational environment (Letier, 2001; van Lamsweerde & Letier, 2000). As mentioned earlier, these situations are referred to as obstacles. They should be systematically identified, assessed, and mitigated against. This may lead to goal model elaboration. If goals are not threatened by any obstacles, the system architect can skip this step and proceed to Step 2 (detailed in Section 3.2). An iterative identify-assess-resolve cycle for the obstacle analysis is required as follows (steps 1.2.1 to 1.2.3).

**Step 1.2.1. Identifying data analytics adoption obstacles.** Obstacles may originate from intrinsic characteristics of data analytics platforms or their operations. The system architect uses domain information to iteratively refine the goal model identifying obstacles and any sub-obstacles (Letier, 2001). In the scenario, the candidate data store technologies for the operationalisation of goal *g7.Achieve [Increased unstructured data storage capacity]* may obstruct the goal *g4.Achieve [Maintained interoperability]* (Figure 4). The reason is that existing ETL's databases are relational. These are not compatible with no-SQL schema-free data storages such as *a17.MongoDB*, *a18.Accumulo*, *a19.HBase*, *a20.Cloudant*, and *a21.BigTable*. Converting thousands of line of complex T-SQL codes defined in ETL to this type of big data storages is not a simple task. In other words, the alternative technologies raise the obstacle *o1.Incompatibility of ETL and big data storages*. This obstacle is further decomposed into sub-obstacles *o1.1.Incompatible datatypes*, *o.1.2.Incompatible data operations*, and *o.1.3.Incompatible APIs*. Data analytics platforms leverage cloud computing servers which are often vulnerable to issues such as bandwidth capacity bottleneck, performance variability or scaling latency, and security (Agrawal, Das, & El Abbadi, 2011). Given that, the partial goal model in Figure 5 represents probable obstacles against the goals that are identified by the system architect.

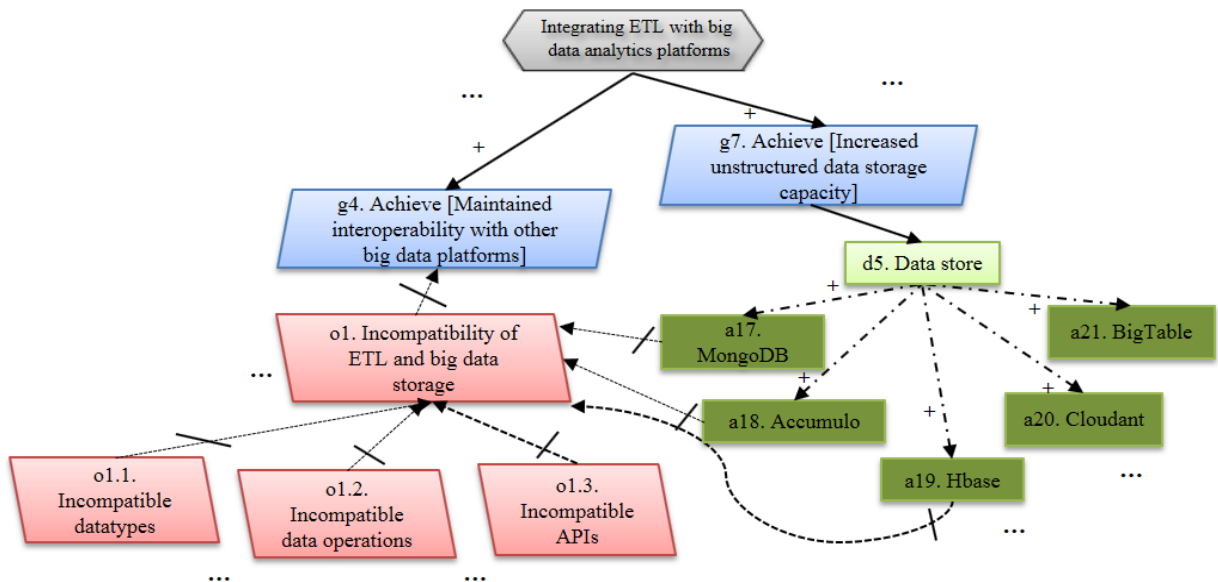

Figure 4. Obstacles to goal *Achieve [Maintained interoperability with other big data platforms]* in the case of using big data store technologies

**Step 1.2.2. Assessing data analytics adoption obstacles.** The identified obstacles from Step 1.2.1 are assessed to generate a new set of architectural requirements. The criticality of the obstacles is judged based on their impact on the goals. Qualitative and quantitative techniques can be employed to perform this step. However, our approach employs a common qualitative technique, Risk Analysis Matrix (Franklin, 1996). This technique specifies the likelihood of an obstacle using a qualitative scale ranging from Almost Certain, Likely, Possible, Unlikely, and Rare. It also indicates the obstacle consequence as Insignificant, Minor, Moderate, Major, and Catastrophic. The risk of an obstacle is defined as the product of its occurrence and severity, i.e. Risk = Likelihood × Consequences. Estimating this risk relies on the availability of domain information sources such as statistics from manufacturing systems, existing accounts on data analytics platforms, or the system architect's judgement. The system architect may conduct a voting technique involving all stakeholders to assess the probability occurrence and severity of obstacles. A risk matrix highlights the risk zone as shown in Table 2. For instance, the risk of an obstacle might be considered as moderate (M), however, it is still tolerable. Whilst a High and Extreme obstacle may necessitate a countermeasure. The values represented in Table 2 are exemplar values.

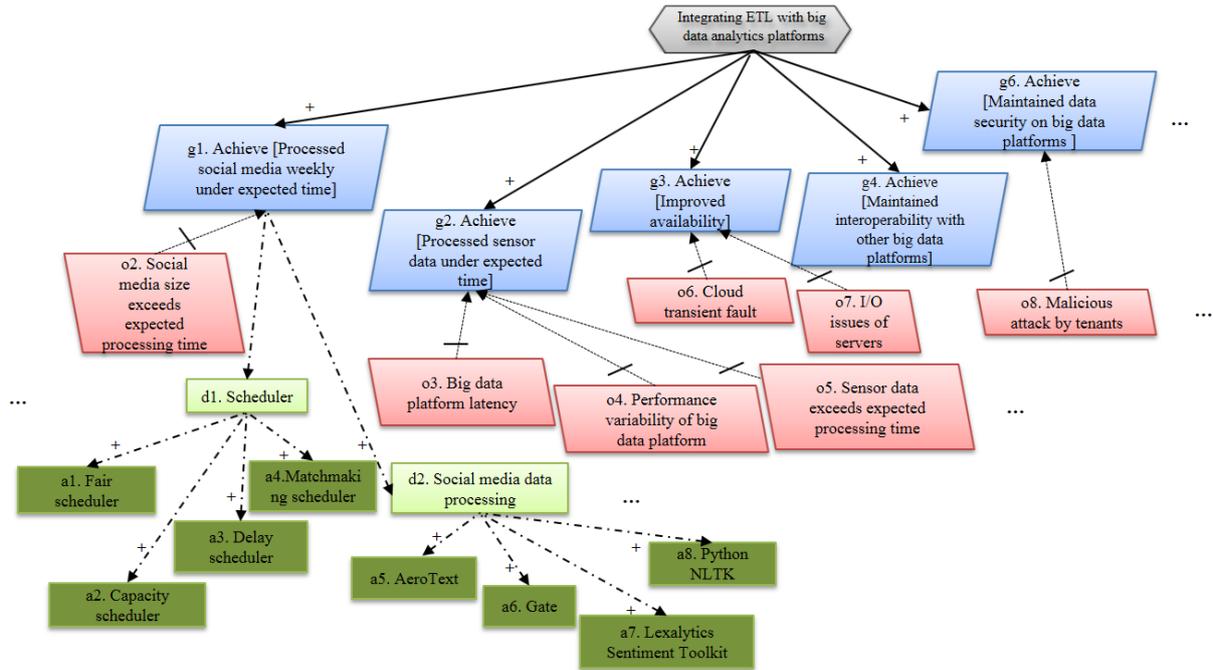

Figure 5. Identified obstacles to goals

Table 2. Risk matrix for obstacle assessment

| Likelihood | Consequence severity | | | | |
|---|---|---|---|---|---|
| | Insignificant | Minor | Moderate | Major | Catastrophic |
| Almost Certain | H | H | E | E | V |
| Likely | M | H | H | E | V |
| Possible | L | M | H | E | E |
| Unlikely | L | L | M | H | E |
| Rare | L | L | M | H | H |

V: Very extreme risk, E: Extreme risk; H: High risk; M: Moderate risk; L: Low risk

**Step 1.2.3 Resolving data analytics adoption obstacles.** Obstacles deemed with severe risk should be resolved. This requires generating new architectural decision alternatives and selecting suitable alternatives amongst them. We employ eight generic and platform independent KAOS's obstacle resolution tactics: *goal substitution*, *agent substitution*, *obstacle prevention*, *goal weakening*, *obstacle reduction*, *goal restoration*, *obstacle mitigation*, and *do-nothing* (Letier & Van Lamsweerde, 2004; van Lamsweerde & Letier, 2000). These are operators on a goal model to refine it to new or existing modified goals, assumptions, and responsibility assignments. They are defined as follows.

**(i) Substitute goal** defines a new alternative goal which is still contributable by data analytics platforms in a way that the obstacle is no longer present. Consider the performance goal *g1.Achieve [Processed social media weekly under expected time]* obstructed by the obstacle *Social media size exceeds processing speed time*. An instance of accommodating this tactic is to collect and process data daily instead of weekly based.

**(ii) Substitute data analytics platform** removes the occurrence of an obstacle by replacing the responsibility for an obstructed goal to a new platform. For example, the obstacle *o5.Sensor data processing exceeds expected time* can be removed via transferring the assigned goal from an overloaded server to another server with lower workload.

**(iii) Prevent obstacle** introduces new assertions to the goal model preventing the obstacle occurrence via applying some factors or doing things in particular way. For instance, consider the security obstacle *o8.Malicious attack by tenants* obstructing the goal *g6.Achieve [Maintained security of sensor data]* (Figure 6). An application of this tactic is to encrypt the batch data collected from sensors prior storing them on big data storages. As such, batch data cannot be read or processed by malicious tenants that are in performing on the same cloud servers. The system architect considers three architectural decision alternatives *o31.Obfuscate data*, *o30.Redact data*, and *o32.Mask data*. Furthermore, to prevent the occurrence of obstacles *o1.1.Incompatible datatypes* and *o.1.2.Incompatible data operations*, architecture decisions *a25.Adapt data* and *a26.Develop adaptor* are considered. It should be noted that employing architecture decision alternatives may cause another set of obstacles against goals. For instance, on the one hand the system architect considers *a25.Adapt data* and *a26.Develop adaptor*. On the other hand, these alternatives may negatively influence the goal *g2.Achieve [Processed sensor data under expected time]*. These dependencies are modelled in Step 2 of the approach described in Section 3.2.

**(iv) Reduce obstacle** introduces agents such as human or servers to behave in certain ways to lessen the occurrence likelihood of an obstacle. Consider the obstacle *o5.Sensor data exceeds expected processing time* to the goal *g2.Achieve [Processed sensor data under expected time]*. An example of applying this tactic is to reduce server workload by prioritising upcoming data that are sent by sensors installed in the product line. The data from highly important sensors that are collected and processed take precedence over those sensors providing supplementary data or do not need a real-time processing. In addition, to reduce the likelihood occurrence of the root obstacle *o4.Performance variability of big data platform*, the architectural decision is *a23.Refine network topology*. Finally, as mentioned earlier, data analytics platforms may be vulnerable to issues such as server latency as represented by the obstacle *o3.Big data analytic platform latency* (Figure 6). To reduce this, the architectural decision *a22.Acquire more resources* (e.g. virtual machines) is chosen.

**(v) Weaken goal** suggests degrading the goal definition to make it more liberal and relaxed in a way that the obstruction no longer occurs. This can be applied in two ways:

-relaxing assumptions of an obstructed goal so that its original form does not needs to be satisfied in all situations. The goal *g2.Achieve [Processed sensor data under expected time]* obstructed by *o5.Sensor data size exceeds expected processing time* is modified to one that the goal is not required to be satisfied in all situations, particularly when data analytics server is not in its fully capacity.

-relaxing the required level of goal satisfaction condition, meaning that there is no further need to full satisfaction. One example of applying this tactic to goal *g2.Achieve [Reduced processing social media weekly under expected time]* is to soften its definition to maximum acceptable time to be achieved by increasing the time/date, i.e. quality variable *processedTime*.

**(vi) Restore goal and mitigate obstacle** are two tactics usable when the avoidance of all obstacles is too costly and tolerating or mitigating consequences of obstacles becomes more practical. In the goal restoration tactic, the system architect adds a new *restoration goal* that prescribes restoration mechanisms for situations when the obstacle impedes the goal. For the obstacle mitigation, the system architect adds a new goal to attenuate the consequences of the obstacle actually occurring. The tactic intends to achieve a weaker satisfaction of an obstructed goal. In the discussed scenario, the goal *g3.Achieve [Improved availability]* is obstructed by the obstacle *o6.Cloud transient fault* as data analytics platforms leveraging cloud might be temporarily unavailable due to reasons such as network traffic or server workload. An implementation technique to mitigate this obstacle is *a27.Retry connection* in the system architecture assuring a weaker version of the goal by specifying next retrying to connect to the server when transient faults occur. Figure 6 shows the goal model after introducing resolution tactics.

**(vii) Do nothing** accepts the risk of an obstacle occurrence.

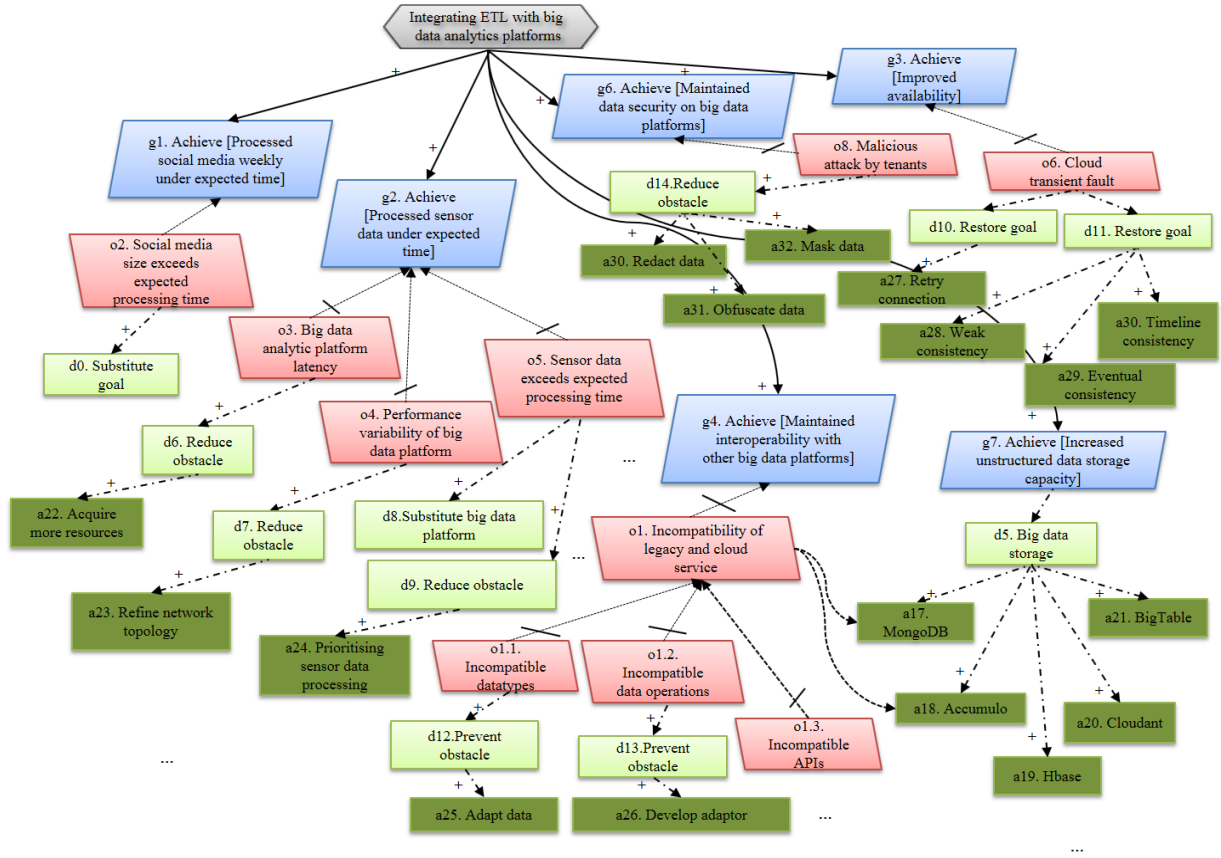

Figure 6. Decision alternatives for handling obstacles

The generated architectural decision alternatives either operationalise goals or tackle obstacles. They form a solution space of different architectures that can be used to integrate ETL with data analytics platforms. Selecting a suitable solution architecture is a challenging task which is systematically dealt in Step 2 of the proposed approach.

As the last note for this step, we believe the scale of an architecture analysis scenario determines whether adopting normative models, like the one presented in this research, necessitates. While an ad-hoc approach for requirement analysis and possible solution architecture is applicable for small-scale projects with a limited number of goals/risks and stakeholder participant, a systematic and communicative presentation layers for specifying notations and model refinements is useful for large-scale projects with multiple goals, potential obstacles, and possible resolution tactics.

## 3.2 Step 2. Exploring uncertainties in big data solution architecture

This step explores candidate solution architecture. The variables that are used in this step presented in Table 3.

Table 3. Symbols used in exploring uncertainty for step 2

| Symbols | Definition |
|---|---|
| $g$ | A goal |
| $G$ | Set of goals |
| $a$ | An alternative to operationalise/implement an architectural decision |
| $A$ | Set of implementation alternatives to operationalise an architectural decision |
| $d$ | A architectural decision |
| $D$ | Set of architectural decisions |
| $x_a$ | If an decision alternative is selected then $x_a$ is 1, otherwise it is 0 |
| $arch$ | A candidate solution architecture including its architectural decision alternatives |
| $AS$ | All possible solution architectures based on all architectural decision alternatives |
| $\widetilde{con}_{g,a}$ | Contribution (positive/negative) of an alternative $a$ on a goal $g$ |
| $\widetilde{S_g}(arch)$ | The total value of a solution architecture with respect to a specific goal $g$ |
| $P_g$ | The weight of the goal $g$ from stakeholders' point of view |
| $Thd_g$ | A threshold to goal $g$ |
| $Thd_c$ | A constraint for the cost of solution architecture |
| $G_{max}$ | A goal $g$ which is expected to be maximised |
| $G_{min}$ | A goal $g$ which is expected to be minimised |
| $\widetilde{cost_a}$ | Fuzzy cost of the alternative $a$ |
| $CH^k(\widetilde{S}(arch_i))$ | Ranking index of solution architecture $i$ with total value $\widetilde{S}$ |

**Step 2.1. Representing uncertain impact of decision alternatives on goals.** We define variable $D$ as a set of architectural decisions. Recall from Step 3.1, the choice for the operationalisation of goal *g7.Achieve [Increased unstructured data storage capacity]* through different big data stores is an example of such a decision (Figure 6). Each architectural decision $d \in D$, itself, may have alternatives for operationalisation, which is specified using set $A_d$. For instance, the decision on *big data storage* has five alternatives, namely: *a17.MongoDB*, *a18.Accumulo*, *a19.HBase*, *a20.Cloudant*, and *a21.BigTable* (Figure 6). As mentioned in Step 3.1, an architectural decision $d$ and its implementation alternatives $A_d$ can be derived using resolution tactics. In the scenario, the architectural decision *prevent obstacle* is used to handle the obstacle *o8.Malicious attack by tenants*. The system architect assumes three different implementation alternatives *a30.Redact data*, *a31.Obfuscate*, and *a32.Mask data* for this architectural decision. This set of implementation alternatives for the decision *prevent obstacle* is defined as $A = \cup_{d \in D} A_d$. The solution space (SS), is a set of all possible alternatives of architectural decisions and their associated implementation techniques. Thus, SA is represented as follows:

$$SS \stackrel{def}{=} \{arch \subseteq A | (\forall d \in D : \exists a \in A_d : a \in arch) \wedge (\forall a \in arch, a \in A_d : \nexists b \in A_d : b \neq a \wedge b \in arch)\}$$

Therefore, with respect to Figure 6, the size of the solution space in the current scenario is 5*5*4*4*3*3*1*1*3 = 10800 potential alternative architectural solutions.

For an implementation alternative $a \in A$ and goal $g \in G$, we define $\widetilde{con_{g,a}}$ which specifies the contribution/impact of the alternative $a$ on the goal $g$. The symbol ~ indicates that the contribution is a fuzzy number. To represent a fuzzy impact, our approaches uses *Triangular fuzzy numbers* (TFNs) (Pedrycz, 1994). TFNs are widely used to represent the approximate value range of linguistic variables. A triangular fuzzy number is represented by A = ($w$, $y$, $z$) where the parameters $w$, $y$, and $z$, respectively, show the smallest possible value, the most promising value, and the largest possible value describing a fuzzy event. TFN linear membership function $\mu A$ is defined by:

$$\mu_A(x) = \begin{cases} \dfrac{x - w}{y - w} & w \leq x \leq y \\ \dfrac{z - x}{z - y} & y \leq x \leq z \\ 0, & otherwise \end{cases}$$

We divide the goal satisfaction into five levels: Very low (VL), Low (L), Medium (M), High (H), and Very high (VH) as shown in Table 4. The numerical range for the goal value and the membership function are respectively presented in Table 5 and Figure 7.

Table 4 Linguistic variables used to show the impact of operationalisation alternatives on quality goals

| Level | Satisfaction value |
|---|---|
| Very low (VL) | 0 and less than 1 |
| Low (L) | Between 0 and 2 |
| Medium (M) | Between 1 and 3 |
| High (H) | Between 2 and 4 |
| Very high (VH) | 3 and more than 3 |

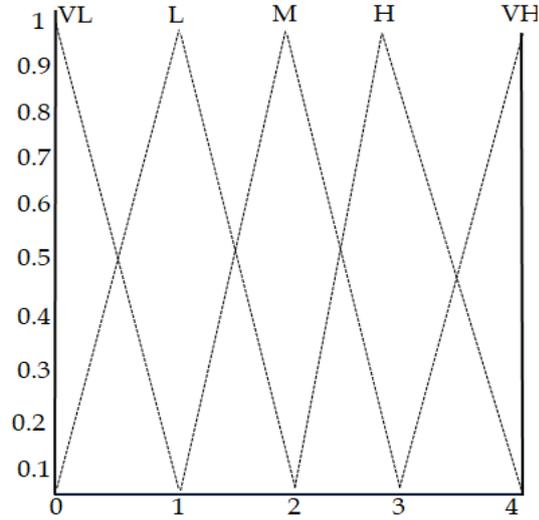

Figure 7. Impact of an implementation alternative on a goal

Table 5. Triangular membership functions for the linguistic variables

$$VH_A(x) = \begin{cases} \frac{x-w}{y-w} & 3 \leq x \leq 4 \\ 1 & 4 \leq x \\ 0, & otherwise \end{cases}$$

$$H_A(x) = \begin{cases} \frac{x-w}{y-w} & 2 \leq x \leq 3 \\ \frac{z-x}{z-y} & 3 \leq x \leq 4 \\ 0, & otherwise \end{cases}$$

$$M_A(x) = \begin{cases} \frac{x-w}{y-w} & 1 \leq x \leq 2 \\ \frac{z-x}{z-y} & 2 \leq x \leq 3 \\ 0, & otherwise \end{cases}$$

$$L_A(x) = \begin{cases} \frac{x-w}{y-w} & 0 \leq x \leq 1 \\ \frac{z-x}{z-y} & 1 \leq x \leq 2 \\ 0, & otherwise \end{cases}$$

$$VL_A(x) = \begin{cases} \frac{x-w}{y-w} & 0 \leq x \leq 1 \\ 0 & x \leq 0 \\ 0, & otherwise \end{cases}$$

For example, given the impact of implementation alternative *a8.Python NLTK*, i.e. a8, on the goal *g2.Achieve [Processed sensor data under expected time]*, i.e. g2, is crisp number 3, the fuzzy representation for the goal satisfaction is:

$$\widetilde{con}_{Achieve\ [Processed\ sensor\ data\ under\ expected\ time], Python\ NLTK} = (0.1, 0.25, 0.25, 0, 0.4)$$

**Step 2.2 Calculating solution architecture value.** This is determined at two levels. Firstly, the fuzzy aggregation of selected implementation alternatives' contributions to a specific goal *g* in a given architecture *arch* ∈ SS is defined through equation (i):

$$\widetilde{S_g}(arch) = \sum_{a\ \in arch}(\widetilde{con}_{g,a} x_a) \quad (i)$$

Note that in (i), for each implementation alternative $a \in A$, we consider a binary decision variable $x_a$, indicating whether an alternative is chosen, i.e. $x_a=1$, or not chosen, i.e. $x_a=0$. To calculate equation (i), we utilise Mamdani fuzzy inference technique (Mamdani, 1974). For example, given the impact of selected 32 operationalisation alternatives on *g2.Achieve [Processed sensor data under expected time]*, the value of a solution architecture in terms of g2 is computed using the fuzzy rules presented in Table 6.

Table 6. An excerpt of fuzzy rules for determining the obtained value for a goal *g2.Achieve [Processed sensor data under expected time]* in a given solution architecture

|    | a1 |     | a2 |     | a3 |     | a4 |     | … | a32 |      | $\widetilde{S_{g2}}$ |
|----|----|-----|----|-----|----|-----|----|-----|---|-----|------|----|
| If | VH | and | VH | and | VH | and |    | and | … | H   | then | H  |
| If | VH | and | VH | and | VH | and |    | and | … | VH  | then | VH |
| If | VH | and | VH | and | VH | and |    | and | … | VH  | then | VH |
| If | VH | and | VH | and | VH | and |    | and | … | VH  | then | VH |
| If | VH | and | VH | and | VH | and |    | and | … | VH  | then | VH |
| …  | …  | …   | …  | …   | …  | …   | …  | …   | … | …   | …    | …  |

The same fuzzy rules are applied for other goals *g2 to g7*. Next, the total value of a given solution architecture *arch* ∈ *SS* is the aggregation of attained fuzzy values for all goals. This is defined via equation (ii):

$$\widetilde{S}(arch) = \sum_{g \in G}(P_g \widetilde{S_g}(arch)) \quad (ii)$$

Again fuzzy rules are defined to determine the total value of a solution architecture based on the fuzzy values of goals as shown in Table 7.

Table 7. An excerpt of fuzzy rules for determining the value of a given solution architecture

|    | **g1** |     | **g2** |     | **g3** |     | **g4** |     | **g5** |     | **g6** |     | **g7** |      | $\widetilde{S}(arch)$ |
|----|----|-----|----|-----|----|-----|----|-----|----|-----|----|-----|----|------|----|
| If | VH | and | VH | and | VH | and | VH | and | VH | and | VH | and | H  | then | H  |
| If | VH | and | VH | and | VH | and | VH | and | VH | and | VH | and | VH | then | VH |
| If | VH | and | VH | and | VH | and | VH | and | VH | and | VH | and | VH | then | VH |
| If | VH | and | VH | and | VH | and | VH | and | VH | and | VH | and | VH | then | VH |
| If | VH | and | VH | and | VH | and | VH | and | VH | and | VH | and | VH | then | VH |
| …  | …  | …   | …  | …   | …  | …   | …  | …   | …  | …   | …  | …   | …  | …    | …  |

The same fuzzy rules are applied for other candidate solution architecture. Our aim is to find an *arch* ∈ *SS* with highest value of $\widetilde{S}(arch)$.

Note that in tables 6 and 7, the number of fuzzy rules can be dramatically increased if the there are many goals, architectural decisions, and operationalisation alternatives. Given the fact that stakeholders may have different emphasises on the quality goals, some fuzzy rules can be removed from tables 6 and 7. In doing so, for each goal $g \in G$, we assign a numeric value between $P_g \in [1..10]$ that shows the degree of priority of the goal in view of stakeholders. As such, for goals with low priority, there is no need to write fuzzy rules.

**Step 2.3. Specifying solution architecture constraints.** A goal may have a certain constraint that has to be satisfied, e.g. the constraint for *g2.Achieve [Processed sensor data under expected time]* is expected to be less than 40 millisecond. A constraint for a goal *g*, represented by $Crt_g$, defined through equation (iii):

$$\forall g \in G_{max} : Crt_g \leq \widetilde{S_g}(arch) \quad (iii)$$
$$\forall g \in G_{min} : \widetilde{S_g}(arch) \leq Crt_g$$

In accepting/rejecting a solution architecture, its cost is also an important constraint, which is shown using $Crt_c$ and defined using equation (iv):

$$\sum_{a \in arch}(\widetilde{cost}_a x_a) \leq Crt_c \quad (iv)$$

$\widetilde{cost}_a$ shows the fuzzy cost of the decision alternative *a*. Constraints on goals and cost are defined regarding project context in which they are applied.

**Step 2.4. Comparing solution architectures.** Given the obtained values for $\widetilde{S}$ (arch), we can compare solution architectures. Between two architectures $arch_1$ and $arch_2 \in SS$, $arch_1$ is more desirable if: $\widetilde{S}(arch_1) \leq \widetilde{S}(arch_2)$. This is a fuzzy comparison of two fuzzy ranges of possible values for two solution architectures resulting in the one with a better range. For this, we employ Chen's method (S.-H. Chen, 1985) where it defines the concepts of fuzzy maximising and minimising sets expressed using equations (v) and (vi):

$$S_{max}(x) = (x - x_{min}/x_{max} - x_{min})^k \quad (v)$$
$$S_{min}(x) = (x_{max} - x/x_{max} - x_{min})^k \quad (vi)$$

In equations (v) and (vi), $S_{max} = \sup \bigcup_{i=1}^{n} \sup S_i$ and $S_{min} = \inf \bigcup_{i=1}^{n} \sup S_i$ and $k > 0$ are real numbers. Using these two sets, left and right utility of a fuzzy number $S_i$, $i^{th}$ architecture solution, is defined as:

$$L(\widetilde{S}(arch_i)) = \sup \min (S_{min}(x), \widetilde{S}(arch_i)) \quad x \in \mathbb{R}$$
$$R(\widetilde{S}(arch_i)) = \sup \min (S_{max}(x), \widetilde{S}(arch_i)) \quad x \in \mathbb{R}$$

Given that, the ranking index for $i^{th}$ solution architecture is obtained using the equation (vii):

$$CH^k(\widetilde{S}(arch_i)) = \frac{1}{2}(R(\widetilde{S}(arch_i)) + 1 - L(\widetilde{S}(arch_i))) \quad (vii)$$

**Step 2.5. Finding the optimum solution architecture.** Finding an architecture optimising quality goals is a typical multi-objective optimisation problem under some constraints (H.-J. Zimmermann, 1978). A solution architecture is considered optimal if it maximises quality goals and satisfies imposed constraints which is, in fact, defined as a linear programming problem through equation (viii):

$$\text{Maximize } \widetilde{S}(arch) \text{ subject to the constraint equations (iii) and (iv)} \quad (viii)$$

(viii) maximises the cumulative value of architectures by selecting a combination of alternatives resulting in the highest architecture value under equations (iii) and (iv) to avoid constraint violation.

## 4 Application exemplar

The scenario of integrating ETL with data analytics platforms (Section 2.1) is used to examine the proposed approach. Tables 8 shows the goals for reengineering that are elaborated into architectural decisions, operationalisation alternatives. They collectively form a space of possible solution architectures for exploration.

Table 8. Goals, architectural decisions, and operationalisation alternatives

| Goal | Architectural decision | Operationalisation alternative |
|---|---|---|
| g1. Achieve [Processed social media weekly under expected time] | d0. Substitute goal | Not applicable (the goal definition is refined) |
| | d1. Social media data processing | a8.Python NLTK |
| | | a6.Gate |
| | | a7.Lexalytics Sentiment Toolkit |
| | | a5.AeroText |
| | d2.Scheduler | a1.Fair scheduler |
| | | a2.Capacity scheduler |
| | | a3.Delay scheduler |
| | | a4.Matchmaking scheduler |
| g2. Achieve [Processed sensor data under expected time] | d3. Real-time stream processing | a9.SQLStream |
| | | a10.Storm |
| | | a11.StreamCloud |
| | d6. Reduce obstacle data analytic platform latency | a22.Acquire more resources |
| | d7. Reduce obstacle Performance variability of data analytics platform | a23.Refine network topology |
| | d8. Substitute data analytics platform | Not applicable |
| | d9. Reduce obstacle sensor data exceeds expected processing time | a24.Prioritizing sensor data processing |
| g3. Achieve [Improved availability] | d10. Restore goal for obstacle cloud transient fault | a27.Retry connection |
| | d11. Restore goal | a29.Eventual Consistency |
| | | a28.Weak Consistency |
| | | a30.Timeline Consistency |
| g4. Achieve [Maintained interoperability with other big data platforms] | d12. Prevent obstacle | a25.Adapt data |
| | d13.Prevent obstacle | a26.Develop adaptor |
| g5. Achieve [Improved data visualisation] | d4. Data visualisation | a16.Google chart |
| | | a15.Tableau |
| | | a14.Data-driven |
| | | a13.document |
| | | a12.Fusion chart |
| g6. Achieve [Maintained data security on big data platform] | d14.Prevent obstacle | a30.Redact data |
| | | a32.Mask data |
| | | a31.Obfuscate data |
| g7. Achieve [Increased unstructured data storage capacity] | d5. Big data storage | a17.MongoDB |
| | | a18.Accumulo |
| | | a19.HBase |
| | | a20.Cloudant |
| | | a21.BigTable |

Through collaboration with the stakeholders, the system architect specifies the impact of operationalisation alternatives on the goals. She used linguistic variables shown in Table 4 and triangular fuzzy membership functions defined in Step 2 (Section 3.2). The complexity of selecting decision alternatives to generate a proper solution architecture in the goal model (Figure 6) is revealed when the system architect is faced with a large number of goals and decision alternatives. The scenario follows 7 goals that are expected to be satisfied. Table 9 shows 32 different operationalisation alternatives in total where for each decision only one alternative can be selected. The numbers in Table 9 are a part of the exemplar of stakeholders reflecting on the impact of decision alternatives on the achievement of the goals. The exemplar is elaborated from (Lin et al., 2016). As mentioned earlier, there are 10800 possible architectural solutions, each of which represents a trade-off among the goals. The numbers are in essence subjective quality measures of the various architectural decisions and are used to illustrate the overall process.

Predicting the impact of potential solution architectures on the goals and finding which portion of the solution space is valid can be a challenging exercise, in particular at the early stage of the data analytics enablement process where the real impact of decision alternatives on quality goals is uncertain. Using the second step of the approach (Section 3.2), the system architect can explore the solution space to find a new suitable architecture for ETL.

Table 9. Impact of operationalisation alternatives on system quality goals (NA: Not applicable)

| Decision | Operationalisation alternative | a) The fuzzy expression of decision alternatives' impact on goals | | | | | | b) Simple crisp expression of decision alternatives' impact on goals (values between 1 and 5) | | | | | |
|---|---|---|---|---|---|---|---|---|---|---|---|---|---|
| | | **Goal** | | | | | | **Goal** | | | | | |
| | | g1 | g2 | g3 | g4 | g5 | g6 | g1 | g2 | g3 | g4 | g5 | g6 |
| d0.Substiute goal | NA | NA | NA | NA | NA | NA | NA | NA | NA | NA | NA | NA | NA |
| | a8.Python NLTK | L | VL | H | L | L | H | 2.7 | 1.8 | 4.3 | 2.4 | 2.2 | 4 |
| | a6.Gate | M | VL | M | M | M | VH | 3.5 | 1.3 | 3.7 | 3.9 | 3.6 | 5 |
| d1. Social media data processing | a7.Lexalytics Sentiment Toolkit | M | M | L | M | L | L | 3.3 | 3.8 | 2.8 | 3.2 | 2.8 | 2 |
| | a5.AeroText | VH | L | H | L | H | L | 5.9 | 2.4 | 4.4 | 2.1 | 4.1 | 2 |
| | a1.Fair scheduler | H | M | VH | M | VH | L | 4.2 | 3.9 | 5 | 3.6 | 5 | 2 |
| d2.Scheduler | a2.Capacity scheduler | VL | VL | M | M | L | L | 1.3 | 1.2 | 3.2 | 3.4 | 2.8 | 2 |
| | a3.Delay scheduler | M | VL | M | M | VH | M | 3.6 | 1.8 | 3.7 | 3.9 | 5 | 3 |
| | a4.Matchmaking scheduler | M | H | H | M | H | L | 3.2 | 4.3 | 4.3 | 3.3 | 4.3 | 2 |
| | a9.SQLStream | VL | H | M | H | L | VH | 1.8 | 4.6 | 3.9 | 4.1 | 2.1 | 5 |
| d3.Real-time stream processing | a10.Storm | M | VL | M | H | L | VL | 3.7 | 1.3 | 3.7 | 4.3 | 2.5 | 1 |
| | a11.StreamCloud | M | M | VH | L | M | L | 3.3 | 3.6 | 5 | 2.7 | 3.2 | 2 |
| | a16.Google chart | L | M | H | M | VH | VL | 2.2 | 3.1 | 4.2 | 3.6 | 5.2 | 1.7 |
| | a15.Tableau | L | M | H | VH | M | VL | 2.3 | 3.7 | 4.6 | 5 | 3.2 | 1.5 |
| d4. Data visualisation | a14.Data-driven | M | M | M | VH | M | VL | 3.2 | 3.8 | 3.7 | 5 | 3.3 | 1.9 |
| | a13.Document | H | M | M | M | H | H | 4.1 | 3.1 | 3.2 | 3.6 | 4.5 | 4.9 |
| | a12.Fusion chart | VL | H | M | M | L | VH | 1.7 | 4.3 | 3.8 | 4.3 | 2.4 | 5 |
| | a17.MongoDB | L | VL | M | H | VH | VH | 2.1 | 1.7 | 3.3 | 4.5 | 5.4 | 5 |
| | a18.Accumulo | L | L | M | M | VH | H | 2.6 | 2.8 | 3.3 | 3.3 | 5 | 4.2 |
| d5. Big data store | a19.HBase | H | L | H | M | VH | VH | 5 | 2.6 | 4.3 | 3.2 | 5 | 5 |
| | a20.Cloudant | M | H | M | VL | M | VL | 3.9 | 4.6 | 3.7 | 1.6 | 3.2 | 1.9 |
| | a21.BigTable | M | M | M | H | H | VH | 3.4 | 3.7 | 3.6 | 4.4 | 4.7 | 5 |
| d6. Reduce obstacle big data analytic platform latency | a22.Acquire more resources | M | M | M | M | VH | H | 3.8 | 3.8 | 3.2 | 3.1 | 5 | 4 |
| d7. Reduce obstacle performance variability of big data platform | a23.Refine network topology | VL | H | M | M | L | VH | 1.7 | 4.3 | 3.8 | 4.3 | 2.4 | 5 |
| d8 – d10 | NA | NA | NA | NA | NA | NA | NA | NA | NA | NA | NA | NA | NA |
| d11. Restore goal | a29.Eventual consistency | VH | M | L | H | VH | H | 5 | 3.3 | 1.7 | 4.7 | 5 | 4 |
| | a28.Weak consistency | VL | M | M | M | M | M | 1.3 | 3.6 | 3.9 | 3.9 | 3.2 | 3 |
| | a30.Timeline consistency | M | VH | M | M | VH | H | 3.8 | 5 | 3.2 | 3.4 | 5 | 4 |
| d12. Prevent obstacle | a25.Adapt data | VL | M | M | H | H | M | 1.4 | 3.9 | 3.5 | 4.5 | 4.8 | 3 |
| d13. Prevent obstacle | a26.Develop adaptor | M | L | VL | M | VL | M | 3.6 | 2.2 | 1.6 | 3.6 | 1.4 | 3 |
| | a30.Redact data | M | VL | L | M | M | M | 3.2 | 1.8 | 2.4 | 3.9 | 3.6 | 3 |
| d14. Prevent obstacle | a32.Mask data | M | M | M | M | VH | H | 3.8 | 3.8 | 3.2 | 3.1 | 5 | 4 |
| | a31.Obfuscate data | VL | H | L | M | M | H | 1.1 | 4.2 | 2.8 | 3.9 | 3.6 | 4 |

Given that the goal weights are assumed to be equal by the stakeholders, the value of each solution architecture is calculated using equation (ii) and fuzzy rules in Table 6. One of the constraints imposed by the stakeholders was to keep the cost of a solution architecture implementation under $30000, which is represented using equation (iv). This constraint ruled out 2531 solution architectures out of 10800. Thus, 8269 solution architectures were left. Following further discussions with the stakeholders, it was agreed to relax the cost constraint to $36000. Subsequently, this reduced the number of rejected architectures to get a better chance for solution exploration. In other words, 564 solution architectures were added, i.e. 8833 solutions in total. As mentioned in Section 3.2, the second constraint imposed by ETL users was to keep the data stream processing coming from sensors below 40 milliseconds. This allows the system architect to assess the system performance constraint on the choice of solution architectures. Relaxing this constraint to 47 milliseconds yielded in increasing the number of acceptable candidate architectures. With this change, 185 solution architectures were further added to the solution space. That is, 9018 valid solutions remained for further analysis. Table 10 shows the top 10 solution architectures ranked using equation (vii) and the selected architectural decision alternatives for all those. Recall from Step 2.4 (Section 3.2), among two solution architectures the one is better if it has a greater fuzzy value. The best solution architecture ranked as the first one has the best combination of decision alternatives in view of trade-off among goals compared to the majority of candidates. That is, it has the best combination of fuzzy values. Nevertheless, it is still likely that the architect may select a solution architecture that is slightly worse than the optimal one due to some reasons (e.g. preference to a particular data analytics platform in the marketplace. Figure 8 shows the final goal model based on the selected decision alternatives for the first ranked solution architecture.

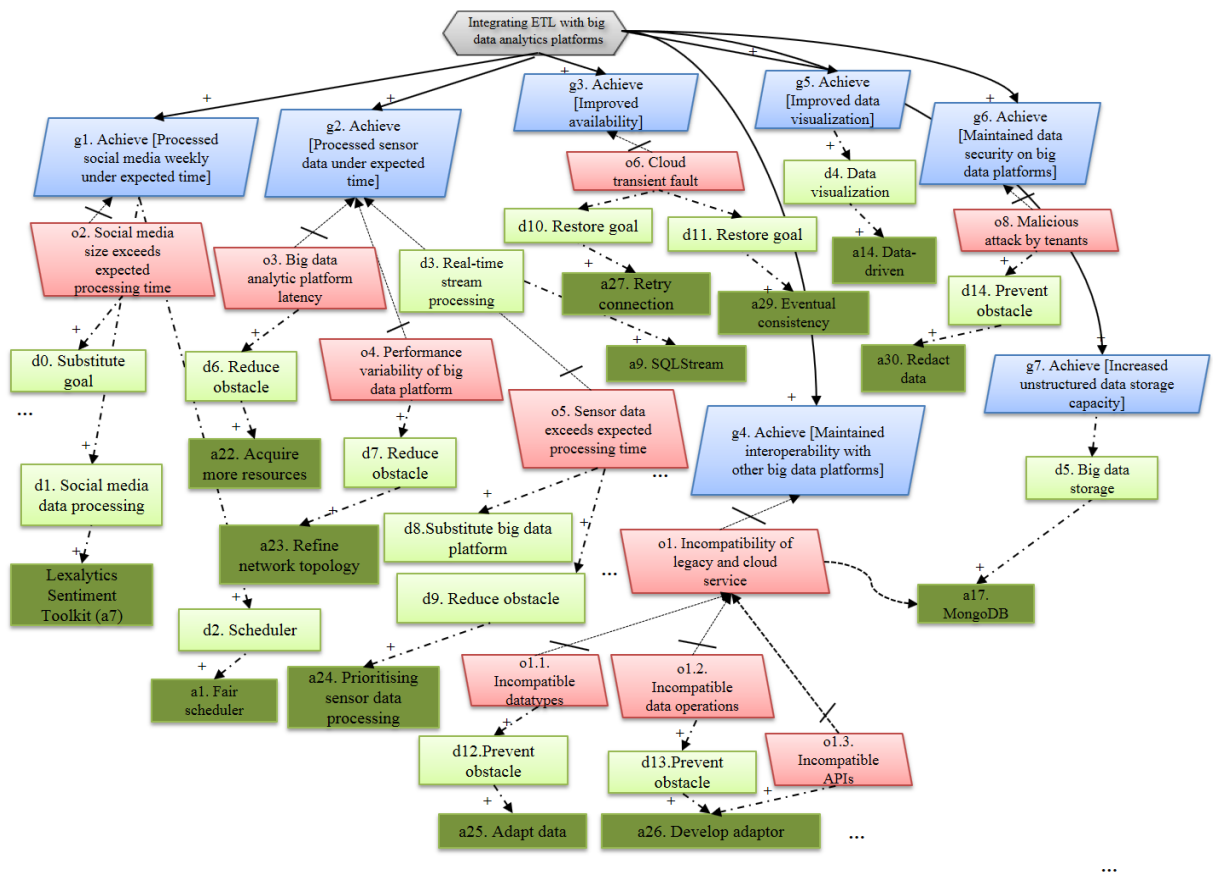

Figure 8. The first solution architecture including the best combination of implementation alternatives

Table 10. Ranked solution architectures with respect to the decision alternatives based on fuzzy-logic and crisp approaches

| | (a) Fuzzy approach | | | | | | | | | | | | | (b) Crisp approach | | | | | | | | | |
|---|---|---|---|---|---|---|---|---|---|---|---|---|---|---|---|---|---|---|---|---|---|---|---|
| | Decision alternative | | | | | | | | | | | | | Decision alternative | | | | | | | | | |
| Rank | d1. Social media data processing | d2. Scheduler | d3. Real-time stream processing | d4. Data visualisation | d5. Data store | d6. Reduce obstacle | d7. Reduce obstacle | d8-d10 | d11. Restore goal | d12. Prevent obstacle | d13. Prevent obstacle | d14. Prevent obstacle | Rank | d1. Social media data processing | d2. Scheduler | d3. Real-time stream processing | d4. Data visualisation | d5. Data store | d6. Reduce obstacle | d7. Reduce obstacle | d8-d10 | d11. Restore goal | d12. Prevent obstacle | d13. Prevent obstacle | d14. Prevent obstacle |
| 1 | Lexalytics Sentiment Toolkit | Fair scheduler | SQLStream | Data-driven | MongoDB | Acquire more resources | Refine network topology | NA | Eventual Consistency | Adapt data | Develop adaptor | Redact data | 1 | Gate | Fair scheduler | Storm | Google chart | Accumulo | Acquire more resources | Refine network topology | NA | Timeline Consistency | Adapt data | Develop adaptor | Obfuscate data |
| 2 | Gate | Capacity scheduler | StreamCloud | Document | Cloudant | Acquire more resources | Refine network topology | NA | Timeline Consistency | Adapt data | Develop adaptor | Mask data | 2 | AeroText | Capacity scheduler | StreamCloud | Document | BigTable | Acquire more resources | Refine network topology | NA | Eventual Consistency | Adapt data | Develop adaptor | Redact data |
| 3 | Gate | Capacity scheduler | StreamCloud | Fusion chart | Cloudant | Acquire more resources | Refine network topology | NA | Timeline Consistency | Adapt data | Develop adaptor | Obfuscate data | 3 | Python NLTK | Matchmaking scheduler | SQLStream | Tableau | Cloudant | Acquire more resources | Refine network topology | NA | Timeline Consistency | Adapt data | Develop adaptor | Mask data |
| 4 | Lexalytics Sentiment Toolkit | Matchmaking scheduler | SQLStream | Data-driven | Accumulo | Acquire more resources | Refine network topology | NA | Weak Consistency | Adapt data | Develop adaptor | Redact data | 4 | Gate | Fair scheduler | StreamCloud | Data-driven | MongoDB | Acquire more resources | Refine network topology | NA | Weak Consistency | Adapt data | Develop adaptor | Mask data |
| 5 | Lexalytics Sentiment Toolkit | Fair scheduler | Storm | Fusion chart | Tableau | Acquire more resources | Refine network topology | NA | Eventual Consistency | Adapt data | Develop adaptor | Mask data | 5 | Lexalytics Sentiment Toolkit | Fair scheduler | SQLStream | Google chart | BigTable | Acquire more resources | Refine network topology | NA | Eventual Consistency | Adapt data | Develop adaptor | Obfuscate data |
| 6 | Gate | Delay scheduler | Storm | Data-driven | Cloudant | Acquire more resources | Refine network topology | NA | Timeline Consistency | Adapt data | Develop adaptor | Obfuscate data | 6 | Gate | Capacity scheduler | Storm | Fusion chart | Cloudant | Acquire more resources | Refine network topology | NA | Weak Consistency | Adapt data | Develop adaptor | Redact data |
| 7 | AeroText | Matchmaking scheduler | StreamCloud | Python NLTK | HBase | Acquire more resources | Refine network topology | NA | Weak Consistency | Adapt data | Develop adaptor | Redact data | 7 | Gate | Delay scheduler | StreamCloud | Document | HBase | Acquire more resources | Refine network topology | NA | Weak Consistency | Adapt data | Develop adaptor | Mask data |
| 8 | Lexalytics Sentiment Toolkit | Capacity scheduler | Storm | Document | Google chart | Acquire more resources | Refine network topology | NA | Weak Consistency | Adapt data | Develop adaptor | Obfuscate data | 8 | AeroText | Capacity scheduler | SQLStream | Tableau | Cloudant | Acquire more resources | Refine network topology | NA | Timeline Consistency | Adapt data | Develop adaptor | Mask data |
| 9 | Gate | Fair scheduler | SQLStream | Fusion chart | MongoDB | Acquire more resources | Refine network topology | NA | Timeline Consistency | Adapt data | Develop adaptor | Redact data | 9 | Lexalytics Sentiment Toolkit | Matchmaking scheduler | StreamCloud | Fusion chart | Cloudant | Acquire more resources | Refine network topology | NA | Timeline Consistency | Adapt data | Develop adaptor | Obfuscate data |
| 10 | AeroText | Delay scheduler | Storm | Python NLTK | BigTable | Acquire more resources | Refine network topology | NA | Eventual Consistency | Adapt data | Develop adaptor | Obfuscate data | 10 | Lexalytics Sentiment Toolkit | Delay scheduler | SQLStream | Data-driven | MongoDB | Acquire more resources | Refine network topology | NA | Eventual Consistency | Adapt data | Develop adaptor | Obfuscate data |

Interestingly, the system architect also compared the results generated through our approach and the simple crisp approach (Table 9-b). Herein, the simple crisp approach refers to an approach that ignores the uncertainty in the impact of decision alternatives on quality goals. Stakeholders used crisp values to represent the impact of decision alternatives on the goals. This difference highlights the contribution of our approach compared to the crisp one in the selection of a proper solution architecture. The simple crisp approach for the calculation the value of a candidate solution architecture would select 125$^{th}$ solution architecture as the optimal solution. This is in contrast to our approach in which 125$^{th}$ approach is ranked as 46$^{th}$ suitable candidate architecture. In other words, 125$^{th}$ has a large negative consequence of uncertainty, which is ignored by the crisp approach. The difference between 1th and 125th solution architectures can be also recognized through specific decision alternatives selected for each architecture. Table 10 represents the selected decision alternatives for the first top ten solution architectures based on two approaches. For example, regarding the information in this right and left sides of the table to find selected decision alternatives, it is observable that our approach chose *a17.MongoDB* for *d5.Big data store* for the first ranked solution architecture whilst the crisp approach chose *a18.Accumulo*.

# 5 Related work

There is a paucity of research focus on the early goal-obstacle analysis and architecture decisions in the scope of integrating manufacturing systems with data analytics platforms. The literature most related is thus subsumed under three research streams: (i) traditional system re-engineering, (ii) reengineering to cloud platforms, and (iii) reengineering to data analytics platforms. Hence, we discuss how our approach presented is positioned in relation to notable research in each stream.

## 5.1 Traditional approaches for legacy system reengineering

Early decision making on selecting solution architectural under uncertainty have been already discussed in the Introduction section. One of the earliest work is by Svahnberg et al. where they provide a multi-criteria decision method using Analytic Hierarchy Process (AHP) supporting comparison of different software architecture candidates for software quality attributes (Svahnberg, Wohlin, Lundberg, & Mattsson, 2003). In its process, two sets of vectors of different candidate architectures with respect to different quality attributes and vice versa are created and refined. The variance of uncertainty is also calculated in each candidate architecture. The sets are used as input for a consensus-based decision making process to identify underlying reasons for disagreements amongst stakeholders. Our proposed equations in this work are inspired by GuideArch approach (Esfahani, Malek, & Razavi, 2013). It presents a fuzzy-based exploration of the architectural solution space under uncertainty aiding architecture in making architecture selection. GuideArch is later extended in (Letier, Stefan, & Barr, 2014) where authors model uncertainty about parameters' values as probability distributions rather than fuzzy values and also assess to extent additional information about uncertain parameters can reduce risks. All of these works including others e.g. (Al-Naeem, Gorton, Babar, Rabhi, & Benatallah, 2005) do not provide a systematic support for top-down goal-obstacle analysis towards generating possible decision architecture alternatives in view of system quality goals. Our approach can be used as a complementary step to generate different alternatives to be used as an input for this group of studies to identify suitable solution architecture.

## 5.2 Legacy systems and cloud computing

An impetus to look at this track of research is the popularity of hosting data solution architectures on the cloud computing platforms (Agrawal et al., 2011). Khajeh-Hosseini et al. (Khajeh-Hosseini, Greenwood, Smith, & Sommerville, 2012) define a cloud adoption conceptual framework to support decision makers in identifying uncertainties. They focus particularly on the cost of deploying options of legacy systems in cloud platforms, which may undergo network latency and service price. Coth studies limit their view to the cost of legacy system reengineering. Similarly, Umar et al. (Umar & Zordan, 2009) defines decision model for reengineering legacy systems to service-oriented architecture to make trade-off between integration versus migration in terms of cost. On the contrary, we do not confine our view to the reengineering cost; rather incorporate other system quality goals that might be important for stakeholders along with elaborating them to potential obstacles and operationalisation alternatives. The approach in (Zardari, Bahsoon, & Ekárt, 2014) uses goal-obstacle analysis to represent risks encountered in using cloud services and mitigating strategies. We extended Zardari's goal-oriented approach by taking into account the risk of uncertainty impact of decision alternatives on stakeholders' goals using fuzzy math. Furthermore, previous cloud migration literature (Alonso, Orue-Echevarria, Escalante, Gorronogoitia, & Presenza, 2013), (Menzel, Schönherr, & Tai, 2013), (Menzel & Ranjan, 2012), and (O. Zimmermann, 2017) suffer providing a meticulous process for an early goal-obstacle exploration and architecture decisions with considering uncertainty issue at the same time.

## 5.3 Legacy systems and big data

At the organisational level, some studies aimed at identifying and analysing business goals for data analytics adoption. For example, Park emphasizes the importance of alignment between organisational business processes and big data sides towards making better business decisions to adopt data analytics platforms (Park, 2017). She defines a systematic process to ensure traceability among high-level big data adoption goals and big data solutions in view of relevance (utility of a data element), comprehensiveness (preventing omissions of potentially important data), and prioritization (required effort in obtaining resources for the data). In another work, Supakkul's approach discusses insights gained from adopting big data to improve business goals (Supakkul, Zhao, & Chung, 2016). Their approach generates two types of resulting insight through goal reasoning and decision-making: (i) descriptive insights of current state of business e.g. the customer retention rate and (ii) predictive insights e.g. customers who are likely to defect. GOBIA (Goal-Oriented Business Intelligence Architecture) is a goal-oriented approach for transforming business goals into a customized big data architecture (Fekete, 2016). GOBIA produces a layered-based conceptual solution architecture, which can be realized by selecting an appropriate mix of data analytics platforms, though it leaves technology selection to implementation phase. The main difference between our approach and the above studies is that we narrow our focus on legacy systems as the unit of analysis and explore solution architectures to integrate them with data analytics platforms.

On the other hand, some work deal with integrating existing legacy systems with data analytics platforms. This genre of literature is deemed closest to our work. A key feature of existing works is their motivations in making legacy systems big data enablement. Some studies develop intelligent techniques such as clustering (Fahad et al., 2014), deep learning (Najafabadi et al., 2015), text mining (Xiang, Schwartz, Gerdes, & Uysal, 2015), and machine learning algorithms (Scott et al., 2016) on big data analytic platforms to mine hiding knowledge in given legacy system data. Once chosen, such techniques can supply inputs to the second step of our approach as decision alternatives for the goal operationalisation or obstacle resolution (see third column of Table 8 for example) where their impact on quality goals is investigated for the optimum selection of solution architecture.

Jha et al. define both forward and backward reengineering activities through which legacy system functionalities are reused, and their data can be accessed and processed by data analytics platforms (S. Jha et al., 2014). They suggest a framework to construct an architectural view of big data solution including business, data, and application architecture (M. Jha, Jha, & O'Brien, 2015). The need for this is highlighted by (Varkhedi, Thati, Nanda, & Alper, 2014) discussing challenges of transferring legacy system data, e.g. mainframes, to a platform configured for big data processing in the same or a separate logical partition on the legacy systems. (Mathew & Pillai, 2015) suggest a three layer-based architecture for handling heterogeneities between legacy systems and data analytics platforms. Govindarajan's work, as a part of Cloud Collaborative Manufacturing Networks (CCMN) project, resolves integrating supply chain manufacturing system and logistic assets with cloud services by developing data adapters for collecting and transforming data from heterogeneous sources to appropriate format accepted by legacy systems, i.e. XML (Govindarajan et al., 2016). Similarly, Givehchi et al. provide a cross-layer architecture enabling interoperability between legacy industrial devices (e.g. I/O devices and sensors) and data analytics platforms (Givehchi, Landsdorf, Simoens, & Colombo, 2017). They apply an information model in order to retain legacy device codes unchanged.

While above approaches acknowledge challenges in legacy system big data enablement scenarios, they keep the description of their analysis process at a high-level that does not represent 'actionable intelligence' towards big data enablement. Nor do they address the uncertainty issue. We have prescribed a more detailed approach for analysing goals in moving manufacturing systems to data analytics platforms, identifying, assessing, and generating resolution tactics in handling potential risks (i.e. step 1 of the approach). Furthermore, we address selecting, prioritization, and ranking resolution tactics in identifying proper solution big data solution architecture under uncertainty (i.e. step 2 of the approach). We have not found other studies that outline a systematic approach on early requirements and big data architecture decisions.

Giret et al. state that the main reason of complexity for developing service-oriented manufacturing systems is the number of heterogeneous technologies and execution environments (Giret, Garcia, & Botti, 2016). They combine multi-agent system design with service-oriented architectures for the development of intelligent automation control and execution of manufacturing systems. Giret later proposes a process model, named Go-green, including activities, guidelines, and tools to design and develop sustainable manufacturing system architectures (Giret, Trentesaux, Salido, Garcia, & Adam, 2017). We believe that the second step of our approach can augment the design phase of Giret' work to fill its gap in addressing early architecture design of big data enabled manufacturing systems under the uncertainty.

# 6 Conclusion, research limitations, and further work

Legacy manufacturing systems are expected to be able to utilize data analytics platforms for advanced information analytics. A clear understanding of goals and risks against data analytics adoption and how they relate to manufacturing systems is particularly crucial. As a business risk management strategy, a systematic architecture design to enable existing manufacturing systems to use data analytics platforms is an important contribution. Our goal-obstacle analysis which takes into account imperfect information and unavoidable uncertainties is quite intuitive to follow. In particular, it provides an early stage analysis, which is taken place before delving into technical aspects of implementing a big data analytics architecture. To the best of our knowledge, such a harness is not available in the literature.

Our approach applies goal reasoning and fuzzy-based logic for analysing suitability of big data solution architecture for manufacturing systems. The approach starts with identifying high-level architectural goals, architectural decision alternatives to realize these goals, generating probable obstacles, and analysing uncertainties in selecting solution architectures. The output of the approach gives the system architect a complete set of architectural requirements to be incorporated into the implementation stage of data analytics architecture implementation to make appropriate trade-offs based on, for instance, cost, security, or performance goals. The application of the approach was also demonstrated the in a scenario of moving ETL to a set of data analytics platforms. Apart from manufacturing and big data settings, due to the genericity of the approach, it can be used in other scenarios of technology adoption when the system architect is interested in evaluating possible solution architecture alternatives.

Our model describes the various options (resolutions) but it does not mandate any. The choice of the resolutions is ultimately determined by whether on the values given to the goals. Hence, strictly speaking, it is not normative per se. However, the scale of complexity involved in the architecture analysis may well lead the architects to rely on the advice produced. Indeed, this is required as we argue. In a lower complexity scenario, the proposed approach may be too intrusive and the system architect may favour an ad-hoc approach for architectural requirements and possible solution architecture. It is important to keep in mind that the work targets settings where systematicity is desirable and actually sought by the system architects. Hence, whilst our approach is applicable for small-scale projects with a limited number of goals/risks and stakeholders, it is more needed for large-scale projects where there are multiple goals, potential obstacles, and possible resolution tactics. Indeed, the approach targets settings where a systematic and communicative approach specifying notations and representation and model refinement mechanisms is useful.

Although we have shown the applicability of our approach, further validation is required to account for the variety in scenarios of integrating manufacturing systems with data analytics platforms. There might be some other ways to satisfy goals or some hidden factors that hinder certain goal achievement but are not defined in the approach's steps. Another important way for the improvement of the approach is to provide further automatic support. The size of the goal model in Step 1 and the number of required computations in Step 2 can limit the usability of the approach without further tool support. We plan to provide a tool support that facilitates using the approach when working with large solution architecture space.

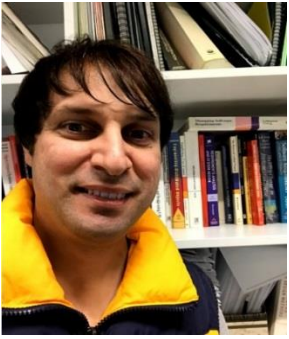

Dr. Mahdi Fahmideh received a PhD degree in Information Systems from the University of New South Wales (UNSW), Sydney Australia. He also holds a master degree in Software Engineering. He is currently a researcher at the University of Technology Sydney. Mahdi is a design science researcher and his vision in research focuses on developing new-to-the-world solutions that help organisations in adopting IT initiatives. His research outputs can be in the form of methodological approaches, frameworks, and conceptual models. Mahdi's research interests lie in the areas of cloud computing, big data, model-driven software development, and situational method engineering. Prior to joining to the academia, Mahdi has worked as a programmer and system analyst in development of software systems in different industry sectors including accounting, insurance, defense, and publishing.

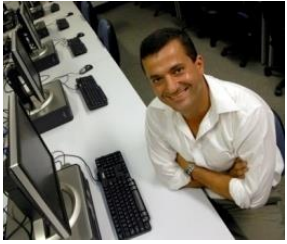

Professor Ghassan Beydoun received a degree in computer science and a PhD degree in knowledge systems from the University of New South Wales. He is currently a Professor of Information Systems at the University of Technology Sydney. He has authored more than 100 papers international journals and conferences. He is currently working on the metamodels for on project sponsored by Australian Research Council and Australian companies to investigate the endowing methodologies for distributed intelligent systems and supply chains with intelligence. His other research interests include multi agent systems applications, ontologies and their applications, and knowledge acquisition.